\documentclass[11pt,twoside,english,preprint]{article}
\usepackage[T1]{fontenc}
\usepackage[latin9]{inputenc}
\usepackage{geometry}
\usepackage{amsmath}
\usepackage{amssymb}
\usepackage{stackrel}
\usepackage{esint}
\PassOptionsToPackage{normalem}{ulem}
\usepackage{ulem}

\makeatletter

\providecommand{\tabularnewline}{\\}

\usepackage{tikz}  
\usepackage{pgfplots}  
 \usetikzlibrary{arrows}

\makeatother

\usepackage{babel}
\begin{document}

\title{On the critical boundary RSOS $\mathcal{M}$$(3,5)$ model }

\author{O. El Deeb}

\maketitle
\begin{center}
Email: o.deeb@bau.edu.lb
\par\end{center}

\begin{center}
Address: Physics Department, Faculty of Science
\par\end{center}

\begin{center}
Beirut Arab University, Beirut, Lebanon
\par\end{center}
\begin{abstract}
We consider the critical non-unitary minimal model ${\cal M}(3,5)$
with integrable boundaries. We analyze the patterns of zeros of the
eigenvalues of the transfer matrix and then determine the spectrum
of the critical theory through the Thermodynamic Bethe Ansatz (TBA)
equations. By solving the TBA functional equation satisfied by the
transfer matrices of the associated $A_{4}$ RSOS lattice model of
Forrester and Baxter in Regime III in the continuum scaling limit,
we derive the integral TBA equations for all excitations in the $(r=1,s=1)$
sector then determine their corresponding energies. The excitations
are classified in terms of $(m,n)$ systems. 
\end{abstract}
keywords: $\mathcal{M}(3,5)$ model, conformal field theory, lattice
models, Yang-Baxter integrability, non-unitary minimal models

\section{Introduction}

Solving 1+1 dimensional Quantum Field Theory (QFT) in finite volume
to determine the energy spectrum and the field correlation functions
is a complicated and non-trivial problem. The volume dependence is
complicated and the energy cannot be calculated exactly even for the
ground state. However, for a special category of models containing
an infinite number of conservation law called integrable models, the
solution is attainable. The bootstrap approach allows the determination
of the masses of the particles and their scattering matrices. The
Bethe-Yang (BY) equations for large volumes are used to determine
an approximate spectrum using the infinite volume quantities \cite{Y,YY}.
This finite size spectrum neglects small vacuum polarization contributions
and contains polynomial corrections inversely proportional to the
volume. Using the Thermodynamic Bethe Ansatz (TBA) method, we can
exactly calculate the vacuum polarization effects for the ground state.
For large Euclidean time, the partition function is dominated by the
ground state contribution. As the roles of space and Euclidean time
can be exchanged by an appropriate transformation, it is sufficient
to evaluate the partition function in the large volume limit, where
Bethe-Yang equations are accurate. Integral equations (TBA) can be
derived for the pseudo energies by calculating the partition function
in the saddle point approximation. The ground-state energy is obtained
from the solutions of these nonlinear TBA equations \cite{Zam1,Zam2,Zam21,Zam22}. 

Extending the TBA method of exploiting the invariance properties of
the partition function to all excited states is not generally an easy
task. However, some information about certain excited states can be
obtained from the exact ground-state TBA equations \cite{DoTa} because
these excited states and the ground-state are related by analytic
continuation. This method using analytic continuation has not been
successfully carried out to obtain TBA equations for the complete
excitation spectrum of the non-unitary $\mathcal{M}(3,5)$ model,
and not even for the simpler non-unitary scaling Lee-Yang model \cite{BajDeeb,Lee}.

Solving the functional relations obtained from the Yang-Baxter integrable
lattice regularization is a powerful and systematic way for obtaining
the TBA integral equations for excited states \cite{PeKlum,KlumPe1,KlumPE,BaxBook}.
These functional equations take the form of $T$ and $Y$ fusion systems.
The $Y$ -system contains the pseudo energies and describes the conformal
spectra at criticality. The functional relations can be recast into
TBA integral equations for the full excitation spectrum using analytic
and asymptotic properties. This method has been successfully applied
for the tricritical Ising model $\mathcal{M}(4;5)$ with conformal
boundary conditions \cite{OBW}. In \cite{PCA1,PCA2}, the tricritical
Ising model was described on the interval and its full spectrum was
determined. The ground-state equation of the sine-Gordon theory and
some excited states were derived in \cite{DesV1,DesV2,FMQR}. The
lattice regularization approach has been systematically developed
for the Lee-Yang theory \cite{BPZ,BDeebP,Deeb}. On the other hand,
the ground-state energy of the Lee-Yang model on the interval and
the excited state TBA equations were derived in \cite{Leclair,DPTW,DTRW}
and defect ground state TBA equations were obtained in \cite{BajSi}. 

The integrable structure of the Conformal Field Theory (CFT) is reflected
in the functional form of the $Y$ -system. In the lattice approach,
it is obtained from the continuum scaling limit of an integrable lattice
regularization of the theory. The lattice approach provides the asymptotic
and analyticity properties and allows the complete classification
of excited states of the theory. 

Our aim is to start developing the next simplest non-unitary model,
namely the critical lattice $\mathcal{M}(3,5)$ model with integrable
boundary conditions. In the present paper we study the critical TBA
equations of the boundary model using a lattice approach. The paper
is organized as follows: In Section 2, we introduce the conformal
model as well as as the continuum scaling limit of the $A_{4}$ RSOS
lattice model of Forrester-Baxter \cite{Huse,Riggs,ABF,ForBax} in
Regime III with crossing parameter $\lambda=\frac{2\pi}{5}$ . We
introduce the commuting double row transfer matrices with integrable
boundaries. We show that the double row transfer matrix satisfies
the universal functional relation in the form of a $Y$ -system after
an appropriate normalization. Section 3 analyzes the conformal spectra
of the transfer matrices. We investigate the analytic structure of
the transfer matrix eigenvalues, classify all excited states of the
trigonometric theory in the $(m,n)$ system and plot representative
zero configurations of an eigenvalue of the transfer matrix. In Section
4, we use the analytic information together with the functional relations
to derive integral TBA equations for the finite volume spectrum of
the $(1,1)$ boundary condition in the critical case, and calculate
the finite-size energies. Lastly, we conclude with discussions in
Section 6.

\section{The $\mathcal{M}(3,5)$ Lattice Model}

The lattice $\mathcal{M}(3,5)$ model under consideration is a Restricted
Solid-on-Solid (RSOS) model defined on a square lattice with heights
that live on an $A_{4}$ Dynkin diagram, with nearest neighbor heights
differing by $\pm1$. It belongs to the general $A_{4}$Forrester-Baxter
models developed in \cite{ABF,ForBax,FPR}

The Boltzmann weights of the general $A_{L}$ Forrester-Baxter models
are given by

\[
\begin{array}{c}
W\left(\begin{array}{cc}
a\pm1 & a\\
a & a\mp1
\end{array}\right)=\frac{s(\lambda-u)}{s(\lambda)}\end{array}
\]

\begin{equation}
W\left(\begin{array}{cc}
a & a\pm1\\
a\mp1 & a
\end{array}\right)=\frac{g_{a\mp1}}{g_{a\pm1}}\frac{s((a\pm1)\lambda)}{s(a\lambda)}\frac{s(u)}{s(\lambda)}
\end{equation}

\[
W\left(\begin{array}{cc}
a & a\pm1\\
a\pm1 & a
\end{array}\right)=\frac{s(a\lambda\pm u)}{s(a\lambda)}
\]

where $a=1,...,L$ , while $u$ is the spectral parameter and $s(u)=\vartheta_{1}(u,p)$
for the massive theory with 

\begin{equation}
\vartheta_{1}(u,q)=2q^{\frac{1}{4}}\sin u\prod_{n=1}^{\infty}\left(1-2q^{2n}\cos2u+q^{4n}\right)\left(1-q^{2n}\right)
\end{equation}

$\vartheta_{1}$ is the elliptic theta function \cite{Gradsh} where
$q$ is the elliptic nome related to the a temperature-like quantity
$T=q^{2}$corresponding to the massive bulk perturbation of the model.
At criticality, $s(u)=\sin(u)$ and corresponds to the conformal massless
model.

The crossing parameter $\lambda$ is given by

\begin{equation}
\lambda=\frac{(p'-p)\pi}{p'}
\end{equation}
 where $p'=L+1$ and $p,p'$ are coprime integers with $p<p'$.

These local face weights satisfy the Yang-Baxter equation which ensures
the integrability of the model. The gauge factors $g_{a}$ are arbitrary
and can be all taken to be equal to $1.$ Unitary models with $p'=p+1$
have positive Boltzmann weights while the non-unitary models with
$p'\neq p+1$ may have negative Boltzmann weights. 

The critical Forrester-Baxter models in Regime III in the continuum
scaling limit

\begin{equation}
\mbox{Regime III:}\ \ \ \ \ 0<u<\lambda,\ \ \ 0<q<1
\end{equation}
 correspond to the minimal models $\mathcal{M}(p,p')$ whose central
charge is 
\begin{equation}
c=1-\frac{6(p-p')^{2}}{pp'}
\end{equation}

Here we consider the $\mathcal{M}(3,5)$ model having $\lambda=\frac{2\pi}{5}$
and $c=-\frac{3}{5}$

A minimal $\mathcal{M}(p,p')$ model has $\frac{(p-1)(p'-1)}{2}$
scaling fields which result in four independent scaling fields for
the $\mathcal{M}(3,5)$ model. As generally prescribed in \cite{CFT},
we can determine the scaling fields, scaling dimensions and fusion
rules. Those fields and their symbols are given in Table 1 below:

\begin{table}[h]
\centering{}\protect\caption{A summary of the different sectors and dimensions of the $\mathcal{M}(3,5)$
model}
\begin{tabular}{|c|c|c|c|}
\hline 
$(r,s)$ & equivalent $\ensuremath{(r,s)}$ & Dimension $\ensuremath{h_{r,s}}$ & Symbol\tabularnewline
\hline 
\hline 
$(1,1)$ & $(2,4)$ & $0$ & $I$\tabularnewline
\hline 
$(2,1)$ & $(1,4)$ & $\frac{3}{4}$ & $\sigma''$\tabularnewline
\hline 
$(2,2)$ & $(1,3)$ & $\frac{1}{5}$ & $\sigma'$\tabularnewline
\hline 
$(2,3)$ & $(1,2)$ & $-\frac{1}{20}$ & $\sigma$\tabularnewline
\hline 
\end{tabular}
\end{table}

The fusion rules of those fields can be obtained using the general
fusion relation

\begin{equation}
\phi_{(r,s)}\times\phi_{(m,n)}=\sum_{\begin{array}{cc}
k=1+|r-m|\\
k+r+m=1\ mod\ 2
\end{array}}^{k_{max}}\ \sum_{\begin{array}{cc}
l=1+|s-n|\\
l+s+n=1\ mod\ 2
\end{array}}^{l_{max}}\phi(k,l)
\end{equation}

where 

\begin{equation}
\begin{array}{cc}
k_{max}=\min(r+m-1,\ 2p'-1-r-m)\\
l_{max}=\min(s+n-1,\ 2p-1-s-n)
\end{array}
\end{equation}

and $k$ and $l$ are incremented by 2. We summarize the fusion rules
of the fields of $\mathcal{M}(3,5)$ here:

\begin{equation}
\begin{cases}
\sigma\times\sigma & =I+\sigma'\\
\sigma\times\sigma' & =\sigma+\sigma'\\
\sigma\times\sigma'' & =\sigma'\\
\sigma'\times\sigma' & =I+\sigma'\\
\sigma'\times\sigma'' & =\sigma\\
\sigma''\times\sigma'' & =I
\end{cases}
\end{equation}

Minimal models have the following fractional decompositions 

\begin{equation}
\frac{p'}{p}=\nu_{0}+1+\frac{1}{\nu_{1}+\frac{1}{\nu_{2}+...+\frac{1}{\nu_{n}+2}}}\ \ \mbox{if \ensuremath{2<2p<p'}}
\end{equation}
and 

\begin{equation}
\frac{p'}{p'-p}=\nu_{0}+1+\frac{1}{\nu_{1}+\frac{1}{\nu_{2}+...+\frac{1}{\nu_{n}+2}}}\ \ \mbox{if \ensuremath{2p>p'}}
\end{equation}

where the parameters satisfy $\nu_{0}>0$ and $\nu_{j}\geq1$ for
$j=1,2,...,n$. and the number of particles in the theory is given
by 
\begin{equation}
t=\sum_{j=0}^{n}\nu_{j}
\end{equation}

In this particular model, with $2p>p',$ we obtain $\nu_{0}=1$ and
all other $\nu_{n\neq0}=0.$ Thus $t=1$ and this model has one type
of particles. This is in direct analogy with its dual $\mathcal{M}(2,5)$
Lee Yang model which only has one type of particles and same values
of $\nu_{n}$.

\subsection{Transfer matrices}

The transfer matrices are constructed from the local face weights.
They form commuting families $[\mathbf{D}(u),\mathbf{D}(v)]=0$ since
the local face weights satisfy the Yang-Baxter equations. This model
satisfies the same functional relation satisfied by the tricritical
hard squares, hard hexagon models and the Lee-Yang model, with spectral
parameter $\lambda=\frac{2\pi}{5}$ instead of $\lambda=\frac{\pi}{5}$
and $\frac{3\pi}{5}$ in the other models \cite{BaxBook,BaxHH,BaxHH1,BaxHS,BDeebP,Deeb}
The new crossing parameter leads to similar analyticity properties
with the Lee-Yang model with a single analyticity strip, but quite
different from the other two aforementioned models having two analyticity
strips.

The double row transfer matrices satisfy the functional relation given
by

\begin{equation}
\mathbf{D}(u)\mathbf{D}(u+\lambda)=1+\mathbf{Y}.\mathbf{D}(u+3\lambda)\label{eq:YY}
\end{equation}

where $\mathbf{Y}$ appearing in \eqref{eq:YY} is the $\mathbb{Z}_{2}$
height reversal symmetry. 

The conformal spectrum of energies $E_{n}$ of the $\mathcal{M}(3,5)$
model can be obtained from the logarithm of the double row transfer
matrix eigenvalue through finite size corrections \cite{Finite-Size}.
In the boundary case, those finite size corrections are given by 

\[
-\log T(u)=Nf_{\mbox{bulk}}(u)+f_{\mbox{boundary}}(u,\xi)+\frac{2\pi}{N}E_{n}\sin\vartheta
\]

where $T(u)$ are the eigenvalues of $\mathbf{D}(u)$ and $N$ is
the number of face weights ($N$ is even in the boundary case) and 

\begin{equation}
\vartheta=\frac{\pi u}{\lambda}=\frac{5u}{2}
\end{equation}
 is the anisotropy angle. 

The bulk free energy and the boundary free energy are given by $f_{\mbox{bulk }}$
and $f_{\mbox{boundary}}$ respectively. Using the inversion relation
methods one can calculate those free energies \cite{energies1,energies2,energies3}

\subsubsection{Boundary weights}

The integrability of this model in presence of a boundary requires
commuting row transfer matrices and triangle boundary conditions that
satisfy the left and right boundary Yang Baxter equations \cite{boundaryint}.
In this model, we label the conformal boundary conditions by the Kac
labels $(r,s)$ where $1\leq r\leq2$ and $1\leq s\leq4$. However,
due to height reversal symmetry it is sufficient to determine the
triangle weights corresponding to independent $(r,s)$ Kac labels
shown in Table 1. These conformal boundaries can be expressed in terms
of integrable boundary conditions in several weights due to gauge
transformations. In fact, as can be proved in the solution of the
boundary Yang-Baxter equations, the $(1,1)$ triangle boundary weights
are arbitrary and here they are given by

\begin{equation}
K_{L}\left(\begin{array}{c}
1\\
1
\end{array}2\biggr|u\right)=\frac{s(2\lambda)}{s(\lambda)},\ \ \ \ \ K_{R}\left(2\begin{array}{c}
1\\
1
\end{array}\biggr|u\right)=1
\end{equation}

The other integrable boundary conditions can be constructed by the
repeated action of a seam on the integrable $(1,1)$ boundary \cite{boundaryfusion}.
The non-zero left and right boundary weights are explicitly calculated
as

\[
K_{L}\left(\begin{array}{c}
2\\
2
\end{array}1\biggr|u,\xi_{L}\right)=W\left(\begin{array}{cc}
2 & 1\\
1 & 2
\end{array}\biggr|u+\xi\right)W\left(\begin{array}{cc}
1 & 2\\
2 & 1
\end{array}\biggr|\lambda-u+\xi\right)K_{L}\left(\begin{array}{c}
1\\
1
\end{array}2\biggr|u\right)
\]

\begin{equation}
=\frac{s(u-2\lambda+\xi_{L})s(u-2\lambda-\xi_{L})}{s(\lambda)^{2}}
\end{equation}

\[
K_{L}\left(\begin{array}{c}
2\\
2
\end{array}3\biggr|u,\xi_{L}\right)=W\left(\begin{array}{cc}
2 & 3\\
1 & 2
\end{array}\biggr|u+\xi\right)W\left(\begin{array}{cc}
1 & 2\\
2 & 3
\end{array}\biggr|\lambda-u+\xi\right)K_{L}\left(\begin{array}{c}
1\\
1
\end{array}2\biggr|u\right)
\]

\begin{equation}
=\frac{s(3\lambda)s(u+\xi_{L})s(u-\xi_{L})}{s(\lambda)^{3}}
\end{equation}

in short notation, we can express this non-zero left boundary weight
by

\begin{equation}
K_{L}\left(\begin{array}{c}
2\\
2
\end{array}a\biggr|u,\xi_{L}\right)=\frac{s(a\lambda)s(u+\xi_{L}+(a-3)\lambda)s(u-\xi_{L}+(a-3)\lambda)}{s(\lambda)^{3}}\ \ \ a=1,3
\end{equation}

and similarly the non-zero right boundary weight by 

\[
K_{R}\left(a\begin{array}{c}
2\\
2
\end{array}\biggr|u,\xi_{R}\right)=W\left(\begin{array}{cc}
a & 2\\
2 & 1
\end{array}\biggr|u+\xi_{R}\right)W\left(\begin{array}{cc}
2 & 1\\
a & 2
\end{array}\biggr|\lambda-u+\xi_{R}\right)K_{R}\left(2\begin{array}{c}
1\\
1
\end{array}\biggr|u\right)
\]

\begin{equation}
=\frac{s(u+\xi_{R}+(2-a)\lambda)s(u-\xi_{R}+(2-a)\lambda)}{s(\lambda)s(2\lambda)}\ \ \ \ a=1,3
\end{equation}

Varying the imaginary parts of $\xi_{L}$ and $\xi_{R}$, one can
obtain different $(r,s)$ conformal boundary conditions in this theory.
The fact that the boundary weights satisfy the left and right boundary
Yang-Baxter equations ensures the integrability of the model in presence
of those boundaries.

\subsubsection{Double row transfer matrix}

The face and triangle boundary weights defined before are used to
construct a family of commuting double row transfer matrices $\mathbf{D}(u)$
\cite{boundaryint}. For a lattice of width $N$, transfer matrix
$\mathbf{D}(u)$ is given by

\[
\mathbf{D}(u)_{\mathbf{a}}^{\mathbf{b}}=\stackrel[c_{0},..,c_{N}]{}{\sum}K_{L}\Bigl(\begin{array}{c}
r\\
r
\end{array}c_{0}\Bigr|\lambda-u\Bigr)W\left(\begin{array}{cc}
r & b_{1}\\
c_{0} & c_{1}
\end{array}\biggr|\lambda-u\right)W\left(\begin{array}{cc}
b_{1} & b_{2}\\
c_{1} & c_{2}
\end{array}\biggr|\lambda-u\right)...W\left(\begin{array}{cc}
b_{N-1} & s\\
c_{N-1} & c_{N}
\end{array}\biggr|\lambda-u\right)
\]

\begin{equation}
\ \ \ \times W\left(\begin{array}{cc}
c_{0} & c_{1}\\
r & a_{1}
\end{array}\biggr|u\right)W\left(\begin{array}{cc}
c_{1} & c_{2}\\
a_{1} & a_{2}
\end{array}\biggr|u\right)....W\left(\begin{array}{cc}
c_{N-1} & c_{N}\\
a_{N-1} & s
\end{array}\biggr|u\right)K_{R}\left(c_{N}\begin{array}{c}
s\\
s
\end{array}\biggr|u\right)
\end{equation}

This matrix satisfies periodicity $\mathbf{D}(u+\pi)=\mathbf{D}(u)$,
commutativity $[\mathbf{D}(u),\mathbf{D}(v)]=0$ and the crossing
symmetry property $\mbox{\ensuremath{\mathbf{D}}}(u)=\mathbf{D}(\lambda-u)$.
In general, $\mathbf{D}(u)$ is not symmetric or normal, but it can
be diagonalized because $\tilde{\mathbf{D}}(u)=\mathbf{G}\mathbf{D(u)}=\tilde{\mathbf{D}}(u)^{T}$
is symmetric where the diagonal matrix $\mathbf{G}$ is expressed
by

\begin{equation}
\mathbf{G}_{\mathbf{a}}^{\mathbf{b}}=\prod_{j=1}^{N-1}G(a_{j},a_{j+1})\delta(a_{j},b_{j})\ \ \ \ \ \mbox{with\ \ \ \ \ \ensuremath{G(a,b)=\begin{cases}
\begin{array}{cc}
\frac{s(\lambda)}{s(2\lambda)}, & \ \ b=1,4\\
1 & \mbox{otherwise}
\end{array}\end{cases}}}
\end{equation}

The normalized transfer matrix is defined by

\begin{equation}
\mathbf{D}(u)=S_{b}(u)\frac{s^{2}(2u-\lambda)}{s(2u+\lambda)s(2u-3\lambda)}\left(\frac{s(\lambda)s(u+2\lambda)}{s(u+\lambda)s(u+3\lambda)}\right)^{N}\mathbf{T}(u)
\end{equation}
In the following analysis we limit our discussion to the $(1,1)$
left and right boundary weights corresponding to the $(r,s)=(1,1)$
boundary. The boundary row transfer matrix reduces to

\[
\begin{array}{cc}
\mathbf{D}(u)_{\mathbf{a}}^{\mathbf{b}}=\stackrel[c_{1},..,c_{N-1}]{}{\sum}K_{L}\left(\begin{array}{c}
1\\
1
\end{array}2\biggr|\lambda-u\right)W\left(\begin{array}{cc}
1 & b_{1}\\
2 & c_{1}
\end{array}\biggr|\lambda-u\right)W\left(\begin{array}{cc}
b_{1} & b_{2}\\
c_{1} & c_{2}
\end{array}\biggr|\lambda-u\right)...W\left(\begin{array}{cc}
b_{N-1} & 1\\
c_{N-1} & 2
\end{array}\biggr|\lambda-u\right)\\
\ \ \ \times W\left(\begin{array}{cc}
2 & c_{1}\\
1 & a_{1}
\end{array}\biggr|u\right)W\left(\begin{array}{cc}
c_{1} & c_{2}\\
a_{1} & a_{2}
\end{array}\biggr|u\right)....W\left(\begin{array}{cc}
c_{N-1} & 2\\
a_{N-1} & 1
\end{array}\biggr|u\right)K_{R}\left(2\begin{array}{c}
1\\
1
\end{array}\biggr|u\right)
\end{array}
\]
and 
\[
S_{b}=1\ \ \ \ \ \mbox{for\ \ensuremath{(r,s)=(1,1)}}
\]

Restricting the analysis to the $Y=+1$ eigenspace, the eigenvalues
of the normalized double row transfer matrix $\mathbf{T}(u)$ satisfies
the universal Y-system independent of the boundary conditions \cite{boundaryint},
hence satisfies the functional equation

\begin{equation}
t(u)t(u+\lambda)=1+t(u+3\lambda)
\end{equation}

\section{Classification of states}

In this section, we analyze the complex zero distributions of the
eigenvalues of the double row transfer matrix, without exploring their
corresponding RSOS paths related to the one-dimensional configurational
sums of Baxter's Corner Transfer Matrices (CTMs) \cite{CTM1,CTM2,CTM3,CTM4},
as our aim is to finally solve the TBA equations of the model. We
consider the behavior of finite excitations above the ground state.

\subsection{(m,n) systems, zero patterns, RSOS paths and characters}

In the critical $\mathcal{M}(3,5)$ lattice model with $\lambda=\frac{2\pi}{5}$,
the face weights and the triangle boundary weights are expressed in
terms of the trigonometric functions $s(u)=\sin(u)$. This model corresponds
to the conformal field theory model with central charge $c=-\frac{3}{5}$
. Its Virasoro algebra has four irreducible modules with characters

\begin{equation}
\chi_{h}(q)=q^{-\frac{c}{24}+h}\sum_{n=0}^{\infty}\dim(V_{n}^{h})q^{n},\ \ \ \ h=0,\frac{1}{5},\frac{3}{4},-\frac{1}{20}
\end{equation}

where $n=E$ is the $L_{0}$ eigenvalue or the energy of the given
state. The eigenvalues are characterized by the location and the pattern
of the zeros in the complex $u-$ plane. The entries of the unrenormalized
transfer matrix are Laurent polynomials in the variables $z=e^{iu}$
and $z^{-1}=e^{-iu}$ of finite degree determined by $N$. The transfer
matrices are commuting families with a common set of $u$-independent
eigenvectors. It follows that the eigenvalues are also Laurent polynomials
of the same degree. Numerical diagonalization gives those polynomials
and numerical factorization gives their zeros. As a result, the eigenvalues
are characterized by the location and the pattern of the zeros in
the complex $u$-plane. We analyze those patterns in terms of the
$(m,n)$ systems. In this paper we analyze the boundary case with
$(r,s)=(1,1)$ boundary.

\subparagraph*{$(m,n)$ systems and zero patterns. }

The single relevant analyticity strip in the complex $u$-plane is
the full periodicity strip

\begin{equation}
\frac{\pi}{5}<\mbox{Re \ensuremath{u}<\ensuremath{\frac{6\pi}{5}}}
\end{equation}

In the boundary case, the transfer matrix is symmetric under complex
conjugation so it is enough to study the eigenvalue zero distributions
on the upper half plane. The zeros form strings and the excitations
are described by the string content in the analyticity strip. Here
we notice the occurrence of four different kinds of strings which
we assign as ``1-strings'', ``short 2-strings'', ``long 2-strings''
and ``real 2-strings''. Figure 1 below gives an example on this
string content for a prototype configuration of zeros.\bigskip{}

\begin{center}
 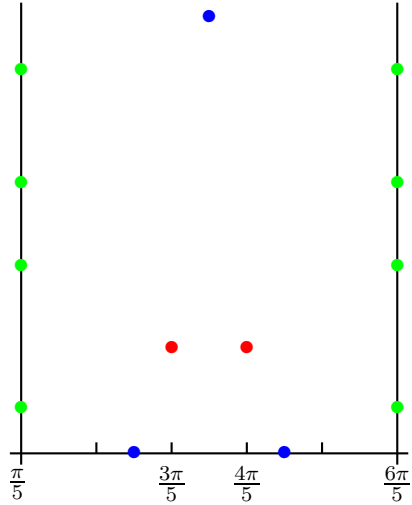
\begin{figure}[h] \begin{center}\begin{tikzpicture}[->,>=stealth',shorten >=0.1 pt,auto,node distance=1cm,   thick,main node/.style={circle,fill=blue!10,draw,font=\sffamily\Large\bfseries}]
\draw[thick,-] (-0.15,0) -- (5.15,0)   ; \draw[thick,-] (0,-0.15) -- (0,6) ; \draw[thick,-] (5,-0.15) -- (5,6); \draw[thick,-] (1,0) -- (1,0.15)   ; \draw[thick,-] (2,0) -- (2,0.15)   ; \draw[thick,-] (3,0) -- (3,0.15)   ; \draw[thick,-] (4,0) -- (4,0.15)   ;
\node [green] at (0,0.6) {\large \textbullet}; \node [red] at (2,1.4) {\large \textbullet}; \node [green] at (0,2.5) {\large \textbullet}; \node [green] at (0,3.6) {\large \textbullet}; \node [green] at (0,5.1) {\large \textbullet};
\node [green] at (5,0.6) {\large \textbullet}; \node [red] at (3,1.4) {\large \textbullet}; \node [green] at (5,2.5) {\large \textbullet}; \node [green] at (5,3.6) {\large \textbullet}; \node [green] at (5,5.1) {\large \textbullet};
\node [blue] at (2.5,5.8) {\large \textbullet}; \node [blue] at (1.5,0) {\large \textbullet}; \node [blue] at (3.5,0) {\large  \textbullet};
 \node [black] at (-0.05,-0.4) { $\frac{\pi}{5}$ };  \node [black] at (2,-0.4) { $\frac{3\pi}{5}$ };  \node [black] at (3,-0.4) { $\frac{4\pi}{5}$ };  \node [black] at (5,-0.4) { $\frac{6\pi}{5}$ };

  \end{tikzpicture}
\caption{A typical configuration of zeros of an eigenvalue of the transfer matrix. The "long 2-string" is in green, the "short 2-string" in red, the "1-string" occurs at the center of the strip furthest from the real axis and the "real 2-string" occurs on the real axis.}
\end{center} \end{figure}
\par\end{center}

\bigskip{}

A 1-string $u_{j}=\frac{7\pi}{10}+iv_{j}$ whose real part is $\frac{7\pi}{10}$
lies in the middle of the analyticity strip. It appears here in the
$(r,s)=(1,1)$ boundary on a fixed location for all eigenvalues. It
appears in some of the sectors/boundaries of the model, while it doesn't
exist in other sectors. Each short 2-string has a pair of zeros whose
real parts are at $\frac{3\pi}{5}$ and $\frac{4\pi}{5}$, with equal
imaginary parts, thus $u_{j}=\frac{3\pi}{5}+iy_{j},\frac{4\pi}{5}+iy_{j}$.
The long 2-string lies at $u_{j}=\frac{\pi}{5}+iy_{j},\frac{6\pi}{5}+iy_{j}$
with equal imaginary parts and with real parts $\frac{\pi}{5}$ and
$\frac{6\pi}{5}.$ The zeroes of a long 2-string lie at the edges
of the analyliticity strip and due to periodicity. In fact, those
2 zeroes are equivalent and correspond to a single zero. The reason
for this naming follows from the general classification of RSOS models
with more than one analyticity strips. Finally, a real 2-string consists
of a pair of zeros $u_{j}=\frac{\pi}{2},\frac{9\pi}{10}$ lying on
the real axis with zero imaginary parts. Due to symmetries, the values
of these real parts are exact for finite $N$

The string contents are described by $(m,n)$ systems \cite{Melzer,Berkovich}
. For this model in the $(1,1)$ sector, we have:

\begin{equation}
2m+n=\frac{N}{2}-2
\end{equation}

where $m$ is the number of short 2-strings, $n$ is the number of
long 2-strings and $N$ is even.

In this sector, we always have a single 1-string furthest from the
real axis, and a real 2-string on the real axis. The 1-string contributes
to one zero, and similarly does the real 2-string due to the symmetry
of the upper and the lower half planes. In addition, each short 2-string
contributes two zeroes, while each long 2-string contributes only
one zero due to periodicity. Hence, the $(m,n)$ system expresses
the conservation of the $2N$ zeroes in the periodicity strip.

Note that the appearance of short 2-strings expresses excited states,
and the absence of short 2-strings occurs in the ground state, where
only long 2-strings appear. For finite excitations, $m$ is finite
while $n\rightarrow N$ as $N\rightarrow\infty.$

An excitation with string content $(m,n)$ is labeled by a unique
set of quantum numbers $I$ 

\[
I=(I_{1},I_{2},....,I_{m})
\]

where the integers $I_{j}\geq0$ equal the number of long 2-strings
whose imaginary parts are greater than that of the given short 2-string
$y_{j}.$ The long 2-strings and short 2-strings labeled by $j=1$
are nearest to the real axis. Those quantum numbers $I_{j}$ satisfy
the equation

\begin{equation}
n\geq I_{1}\geq I_{2}\geq....\geq I_{m}\geq0
\end{equation}

For the example given above in Figure 1, we simply have $I_{1}=3$.
No other quantum numbers $I_{j}$ exist as $m=1$. 

For any string content $(m,n)$, the lowest excitation occurs when
all of the short 2-strings are further away from the real axis than
all of the long 2-strings. This is equivalent to say that all $I_{j}=0$.
Bringing the location of a short 2-string closer to the real axis
below a long 2-string increases its quantum number by one and increases
the energy.

\subsection{Continuum scaling limit}

In the continuum scaling limit, where the even $N\rightarrow\infty,$
the spacing of the zeroes becomes denser. We find that the imaginary
parts of the furthest zeros from the real axis grow as $\frac{2}{5}\log N$
, hence the spacing between zeros tends to $0$ as $\frac{2}{5}\frac{\log N}{N}$.
Finite energy states for large $N$ have zero patterns as depicted
in Figure 2. We denote the imaginary part of the 1-string by $\alpha$
and those of the short 2-strings by $\beta_{j}$. The number of short
2-strings is finite. $\alpha$, $\beta_{j}$, and the imaginary parts
of the long 2-strings furthest from the real axis scale as $\frac{2}{5}\log N$
in the continuum scaling limit.

\bigskip{}

\begin{center}
 \begin{figure}[h] \begin{center}\begin{tikzpicture}[->,>=stealth',shorten >=0.1 pt,auto,node distance=1cm,   thick,main node/.style={circle,fill=blue!10,draw,font=\sffamily\Large\bfseries}]

\draw[thick,-] (-0.15,0) -- (5.15,0)   ; \draw[thick,-] (0,-0.15) -- (0,6) ; \draw[thick,-] (5,-0.15) -- (5,6); \draw[thick,-] (1,0) -- (1,0.15)   ; \draw[thick,-] (2,0) -- (2,0.15)   ; \draw[thick,-] (3,0) -- (3,0.15)   ; \draw[thick,-] (4,0) -- (4,0.15)   ;
\node [green] at (0,4.109) {\large \textbullet}; \node [green] at (0,2.719) {\large \textbullet}; \node [green] at (0,1.906) {\large \textbullet}; \node [green] at (0,1.596) {\large \textbullet}; \node [green] at (0,1.323) {\large \textbullet}; \node [green] at (0,1.092) {\large \textbullet}; \node [green] at (0,0.881) {\large \textbullet}; \node [green] at (0,0.690) {\large \textbullet}; \node [green] at (0,0.515) {\large \textbullet}; \node [green] at (0,0.354) {\large \textbullet}; \node [green] at (0,0.206) {\large \textbullet}; \node [green] at (0,0.067) {\large \textbullet};
\node [blue] at (2.5,5.5) {\large \textbullet}; \node [blue] at (1.5,0) {\large \textbullet}; \node [blue] at (3.5,0) {\large  \textbullet};
\node [red] at (2,3.296) {\large \textbullet}; \node [red] at (3,3.296) {\large \textbullet}; \node [red] at (2,2.271) {\large \textbullet}; \node [red] at (3,2.271) {\large \textbullet};
\node [green] at (5,4.109) {\large \textbullet}; \node [green] at (5,2.719) {\large \textbullet}; \node [green] at (5,1.906) {\large \textbullet}; \node [green] at (5,1.596) {\large \textbullet}; \node [green] at (5,1.323) {\large \textbullet}; \node [green] at (5,1.092) {\large \textbullet}; \node [green] at (5,0.881) {\large \textbullet}; \node [green] at (5,0.690) {\large \textbullet}; \node [green] at (5,0.515) {\large \textbullet}; \node [green] at (5,0.354) {\large \textbullet}; \node [green] at (5,0.206) {\large \textbullet}; \node [green] at (5,0.067) {\large \textbullet};
  \node [black] at (-0.05,-0.4) { $\frac{\pi}{5}$ };  \node [black] at (2,-0.4) { $\frac{3\pi}{5}$ };  \node [black] at (3,-0.4) { $\frac{4\pi}{5}$ };  \node [black] at (5,-0.4) { $\frac{6\pi}{5}$ };

  \end{tikzpicture}
\caption{A typical zero configuration for an eigenvalue in the (1,1) sector for large value of N.}
\end{center} \end{figure}
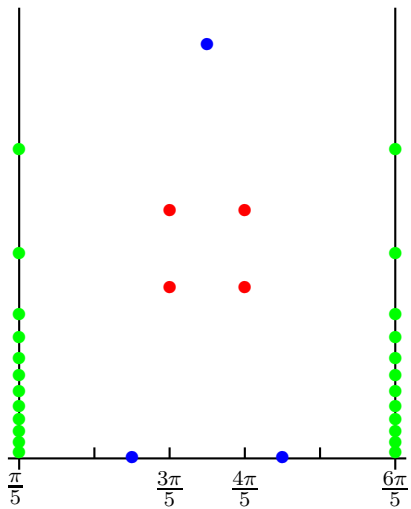
\par\end{center}

\section{Critical TBA Equations}

\subsection{Critical TBA}

The critical TBA equations of this model are derived by solving the
functional relation

\begin{equation}
t(u)t(u+\lambda)=1+t(u+3\lambda)\label{eq:t(u)}
\end{equation}
 while using the analytic structure of the function $t(u)$ which
is factorized according to the large volume behavior as

\begin{equation}
t(u)=f(u)g(u)l(u)
\end{equation}

where $\log f(u)$ is of order $N$, $\log g(u)$ of order 1 and $\log l(u)$
is on the order $\frac{1}{N}$. The leading order term satisfies the
relation 
\[
f(u)f(u+\lambda)=1
\]
 and contains the order $N$ zeros and poles of the normalization
factor. The function $g(u)$ satisfies a similar relation 
\[
g(u)g(u+\lambda)=1
\]
 and accounts for the order 1 boundary dependent zeroes and poles.
The remaining finite size function $l(u)$ is derived from an appropriate
integral equation.

The energy of the states can be calculated from the finite size corrections
of the double row transfer matrix eigenvalues

\begin{equation}
\log t(u)=\log f(u)+\log g(u)-\frac{i}{N}\left(e^{-\frac{5i}{2}u}E^{+}-e^{\frac{5i}{2}u}E^{-}\right)
\end{equation}

where the finite-size conformal energies are given in their exponential
form as

\begin{equation}
E=e^{-\frac{5i}{2}u}E^{+}-e^{\frac{5i}{2}u}E^{-}
\end{equation}

The first term dominates in the $u\rightarrow+i\infty$ limit, while
the second dominates in the $u\rightarrow-i\infty$ 

Using equation \eqref{eq:t(u)}, we find that 

\begin{equation}
t(u)t(u+\frac{2\pi}{5})=1+t(u+\frac{9\pi}{5})
\end{equation}

Exploiting the periodicity of the transfer matrix of $t(u)=t(u+\pi)$
and after an appropriate shift in the variables we obtain the functional
relation

\begin{equation}
t(u-\frac{\pi}{5})t(u+\frac{\pi}{5})=1+t(u)
\end{equation}

The normalization introduces zeros of order $N$ (even) at $\frac{\pi}{5}$
and $\frac{6\pi}{5}$, and poles of order $N$ at $\frac{3\pi}{5}$
and $\frac{4\pi}{5}$. They should be normalized by the function $f(u)$
whose solution compatible with the analytic structure is 
\begin{equation}
f(u)=\left(-\frac{\sin(\frac{5u}{4}-\frac{\pi}{4})\sin(\frac{5u}{4}+\frac{\pi}{2})}{\cos(\frac{5u}{4}-\frac{\pi}{4})\cos(\frac{5u}{4}+\frac{\pi}{2})}\right)^{N}=\left(\frac{\cos(\frac{5u}{2}+\frac{\pi}{4})+\cos\frac{\pi}{4}}{\cos(\frac{5u}{2}+\frac{\pi}{4})-\cos\frac{\pi}{4}}\right)^{N}
\end{equation}

This function satisfies the relation 
\begin{equation}
f(u-\frac{\pi}{5})f(u+\frac{\pi}{5})=1
\end{equation}

In addition to the relation $f(u)f(u+\lambda)=1$ stated above.

In the thermodynamic limit, the imaginary part of the outermost string
from the real $u$ axis goes to infinity as $\frac{2}{5}\log\kappa N$
with 

\begin{equation}
\kappa=4\sin\frac{\pi}{4}=2\sqrt{2}
\end{equation}
 Consequently, we define a real variable $x$ as a vertical coordinate
along the center of the analyticity strip as:

\begin{equation}
u=\frac{7\pi}{10}+\frac{2ix}{5}
\end{equation}

In this new coordinate $x,$ the functional relation becomes

\begin{equation}
t(x-i\frac{\pi}{2})t(x+i\frac{\pi}{2})=1+t(x)\label{eq:functional}
\end{equation}

And upon this change of variable, $f(u)$ becomes 

\begin{equation}
f(x)=\left(\frac{\cosh x+\cos\frac{\pi}{4}}{\cosh x-\cos\frac{\pi}{4}}\right)^{N}
\end{equation}
and satisfies the functional relation 
\begin{equation}
f(x-i\frac{\pi}{2})f(x+i\frac{\pi}{2})=1\label{eq:fx}
\end{equation}

The boundary normalization also introduces a double zero at $u=\frac{\lambda}{2}=\frac{\pi}{5}$
and poles at $u=-\frac{\lambda}{2}+\pi=\frac{4\pi}{5}$ and at $u=\frac{3\lambda}{2}=\frac{3\pi}{5}$.
Due to the presence of the argument $2u$, the periodicity of order
1 functions is $\frac{\pi}{2}$. Thus, we have a double zero at $u=\frac{7\pi}{10}$,
and poles at \uline{$u=\frac{3\pi}{10}$}and $u=\frac{11\pi}{10}$.
Finally, the presence of a real 2-string indicates a couple of zeroes
at $\frac{\pi}{2}$ and $\frac{9\pi}{10}$. 

To account for those zeros and poles, $g(u)$ is defined as 

\[
g(u)=\frac{\tan^{2}(\frac{5u}{4}-\frac{\pi}{4})\tan^{2}(\frac{5u}{4}-\frac{7\pi}{8})\tan(\frac{5u}{4}-\frac{5\pi}{8})\tan(\frac{5u}{4}-\frac{9\pi}{8})}{\tan\frac{5u}{4}\tan(\frac{5u}{4}-\frac{3\pi}{4})}
\]

Transforming into the $x$ variable we obtain

\begin{equation}
g(x)=\frac{\tan^{2}(\frac{ix}{2}+\frac{5\pi}{8})\tan^{2}(\frac{ix}{2})\tan(\frac{ix}{2}+\frac{\pi}{4})\tan(\frac{ix}{2}-\frac{\pi}{4})}{\tan^{2}(\frac{ix}{2}+\frac{7\pi}{8})\tan(\frac{ix}{2}+\frac{\pi}{8})}
\end{equation}

The function $g(u)$ satisfies the functional relations $g(u-\frac{\pi}{5})g(u+\frac{\pi}{5})=1$
which is equivalent to 

\begin{equation}
g(x-i\frac{\pi}{2})g(x+i\frac{\pi}{2})=1\label{eq:gx}
\end{equation}

The order one $g$ term appears in the lattice boundary TBA but it
doesn't explicitly contribute to the energy. Due to the symmetry with
respect to the real $u$-axis obtained by complex conjugation, the
variables are scaled around $\log\kappa N$, and disappear in the
scaling limit.

\subsubsection*{Quantum states}

To find the quantum states and the corresponding critical TBAs, we
need to solve the functional relation \eqref{eq:functional}. To do
this we need to ensure that $l(x)$ is analytic and non-zero in the
analyticity strip, and that its logarithm has constant asymptotic
behavior as $x\rightarrow\pm\infty.$ This is done by characterizing
the eigenvalues of the transfer matrix by their patterns of zeros
in the analyticity strip $\frac{\pi}{5}<u<\frac{6\pi}{5}$. The long
2-strings occur at the boundaries of the analyliticity strip, and
they become dense in the thermodynamic limit $N\rightarrow\infty$,
consequently they define the boundaries of the analyticity strip at
$\frac{\pi}{5}$ and $\frac{5\pi}{6}$. 

In the $(1,1)$ sector, a single 1-string appears at the center of
the strip furthest out from the real axis, with symmetry in the upper
and lower parts of the $u$-plane. Its position always occurs at

\begin{equation}
u_{0}=\frac{7\pi}{10}+i\alpha
\end{equation}
 The short 2-strings correspond to finite excitations above the ground
state and their real parts can occur at $\frac{3\pi}{5}$ and $\frac{4\pi}{5}$,
and they are expressed as

\begin{equation}
u_{j}=\begin{cases}
\frac{3\pi}{5} & +i\beta_{j}\\
\frac{4\pi}{5} & +i\beta_{j}
\end{cases}
\end{equation}

In the thermodynamic limit, with $N\rightarrow\infty$, those zeros
in the scaling regions furthest from the real axis approach infinity
in the upper and lower half planes as

\[
\begin{cases}
\alpha= & \frac{2}{5}\left(\pm\log\kappa N+\tilde{\alpha}^{\mp}\right)\\
\beta= & \frac{2}{5}\left(\pm\log\kappa N+\tilde{\beta}_{j}^{\mp}\right)
\end{cases}
\]

Transforming into the $x=\frac{5}{2i}(u-\frac{7\pi}{10})$ variable,
we find that the locations of the zeros of the 1-strings occur at:

\begin{equation}
x_{0}^{\pm}=\frac{5\alpha}{2}=\pm\log\kappa N+\tilde{\alpha}^{\mp}\label{eq:alphahat}
\end{equation}
 while the zeros of the short 2-strings occur at:

\begin{equation}
\begin{cases}
\left(x_{j}^{\pm}+\frac{i\pi}{4},\ x_{j}^{\pm}-\frac{i\pi}{4}\right)\\
\\
x_{j}^{\pm}=\pm\log\kappa N+\tilde{\beta}_{j}^{\mp}
\end{cases}\label{eq:betahat}
\end{equation}

The remaining task is to convert the functional equation into an integral
TBA equation that can be solved by Fourier transforms in the continuum
scaling limit. To satisfy ANZC functions that are free of zeros and
poles in the strip containing $\mbox{Im}x\in[-\frac{\pi}{2},\frac{\pi}{2}]$,
appropriate functions are introduced to remove the 1-string and the
short 2-string zeros. 
\begin{equation}
\sigma_{0}=\tan\left(\frac{5u}{4}+\frac{\pi}{8}\right)
\end{equation}
 removes the zero of the one string while 
\begin{equation}
\sigma_{1}=-\tan(\frac{5u}{4})\tan(\frac{5u}{4}+\frac{\pi}{4})=\frac{\cos(\frac{5u}{2}+\frac{\pi}{4})-\cos\frac{\pi}{4}}{\cos(\frac{5u}{2}+\frac{\pi}{4})+\cos\frac{\pi}{4}}
\end{equation}
 removes the two zeros of the short two strings.

In the $x$ variable those functions are given by

\begin{equation}
\begin{cases}
\sigma_{0}=\tanh\frac{x}{2}\\
\sigma_{1}=\frac{\cosh x-\cos\frac{\pi}{4}}{\cosh x+\cos\frac{\pi}{4}}
\end{cases}\label{eq:excitations}
\end{equation}

Those functions satisfy the relations

\begin{equation}
\begin{array}{cccc}
\sigma_{0}(x-\frac{i\pi}{2}) & \sigma_{0}(x+\frac{i\pi}{2})=1 & \ ;\  & \sigma_{1}(x-\frac{i\pi}{2})\sigma_{1}(x+\frac{i\pi}{2})=1\end{array}
\end{equation}

Consequently, the appropriate parametrization of the normalized transfer
matrix eigenvalue is

\begin{equation}
t(x)=f(x)g(x)\prod_{\pm}\sigma_{0}(x-x_{0}^{\pm})\prod_{j=1}^{M}\sigma_{1}(x-x_{j}^{\pm})l(x)
\end{equation}

where $M$ is the number of short 2-strings in the strip for a given
excited state.

Using the functional relation \eqref{eq:functional}, and exploiting
the properties of the functions $f(x)$ and $g(x)$ given in \eqref{eq:fx}
and \eqref{eq:gx} , we obtain the equality

\begin{equation}
l(x-i\frac{\pi}{2})l(x+i\frac{\pi}{2})=1+t(x)
\end{equation}

With our construction of all necessary functions, $l(x)$ is analytic
and non-zero in the analyticity strip, and its logarithm has constant
asymptotic (ANZC) as $x\rightarrow\pm\infty$. Taking the logarithm
on both sides and solving the equations using Fourier transforms of
the derivatives $[\log l(x)]'$ we obtain that:

\begin{equation}
\log l(x)=-\varphi\star\log\left[1+t(x)\right]
\end{equation}

where the convolution $\star$ is defined by

\begin{equation}
\left(f\star g\right)(x)=\left(g\star f\right)(x)=\frac{1}{2\pi}\intop_{-\infty}^{+\infty}f(x-y)g(y)dy
\end{equation}

and the function $\varphi$ and its transform $\hat{\varphi}$ are
given by

\begin{equation}
\varphi(x)=\frac{1}{2\pi}\intop_{-\infty}^{+\infty}dk\hat{\varphi}(k)e^{ikx}
\end{equation}

and 

\begin{equation}
\hat{\varphi}(k)=-\frac{1}{e^{k\frac{\pi}{2}}+e^{-k\frac{\pi}{2}}}
\end{equation}

Consequently, and following the procedure of \cite{BDeebP}, an explicit
expression of $\varphi(x)$ can be obtained as:

\begin{equation}
\varphi(x)=-\frac{1}{2\pi\cosh x}
\end{equation}

In general, the kernel $\varphi(x)$ is related to the two-particle
$S$-matrix of the corresponding continuum model, but this $S$-matrix
is not explicitly determined yet.

Restoring $t(x)$ we obtain the critical TBA equations on the lattice
for the $(1,1)$ boundary condition as

\begin{equation}
\log t(x)=\log f(x)+\log g(x)+\sum_{\pm}\log\sigma_{0}(x-x_{0}^{\pm})+\sum_{j=1}^{M}\log\sigma_{1}(x-x_{j}^{\pm})-\varphi\star\log\left[1+t(x)\right]
\end{equation}

The parameters of the excited state $x_{i}=\left\{ x_{0}^{\pm},x_{j}^{\pm}\right\} $
are determined self-consistently from the fact that they are zeros
of the transfer matrix: 

\[
t(x)\biggr|_{x=x_{i}\pm\frac{i\pi}{2}}=-1
\]
In the continuum scaling limit with $N\rightarrow\infty,$ $f(x)$
has nontrivial behavior in the two scaling regions $x\sim\pm\log\kappa N$.
Then, two scaling functions are introduced as

\begin{equation}
e^{\epsilon^{\mp}(x)}=\lim_{N\rightarrow\infty}t(x\pm\log\kappa N)
\end{equation}
 The behavior of $f(x)$ in the scaling regions in important, and
in this scaling limit we obtain that

\begin{equation}
\lim_{N\rightarrow\infty}\log f(x\pm\log\kappa N)=\lim_{N\rightarrow\infty}N\log\left(1+\frac{e^{\mp x}}{N}\right)=e^{\mp x}
\end{equation}

It is interesting to observe that $g(x)$ scales to 1 around $\log\kappa N$,
hence it has no contribution to the subsequent TBA equations, and
no explicit contribution to the energy.

This leads to the massless boundary TBA equations 

\begin{equation}
\epsilon^{\mp}(x)=e^{\mp x}+\sum_{\pm}\log\sigma_{0}(x-\tilde{\alpha}^{\mp})+\sum_{j=1}^{M}\log\sigma_{1}(x-\tilde{\beta_{j}}^{\mp})-\varphi\star\log\left(1+e^{\epsilon^{\mp}(x)}\right)
\end{equation}

The location of the zeros $\tilde{\alpha}^{\mp}$ and $\tilde{\beta_{j}}^{\mp}$
were defined in equations \eqref{eq:alphahat} and \eqref{eq:betahat}.

The lowest energy state of this sector, or what we may call the ground
state of the $(1,1)$ sector has no short strings that represent excitations,
hence the term $\sigma_{1}(x)\rightarrow1$, and does not appear in
the equations so the corresponding massless boundary TBA equation
for the lowest energy state in this sector is given by

\begin{equation}
\epsilon^{\mp}(x)=e^{\mp x}+\sum_{\pm}\log\sigma_{0}(x-\tilde{\alpha}^{\mp})-\varphi\star\log\left(1+e^{\epsilon^{\mp}(x)}\right)
\end{equation}

\subsection{Energies}

The finite size energies of excited states can be determined from
$\log l(x)$ and equations \eqref{eq:excitations} . The energy formula
is 

\[
\frac{1}{N}\left(e^{x}E^{+}+e^{-x}E^{-}\right)=\sum_{\pm}\log\sigma_{0}(x-x_{0}^{\pm})+\sum_{j=1}^{M}\log\sigma_{1}(x-x_{j}^{\pm})-\log l(x)
\]

\begin{equation}
=\sum_{\pm}\log\sigma_{0}(x-x_{0}^{\pm})+\sum_{j=1}^{M}\log\sigma_{1}(x-x_{j}^{\pm})+\intop_{-\infty}^{\infty}\frac{dy}{2\pi}\varphi(x-y)\log\left(1+e^{\epsilon^{\mp}(y)})\right)
\end{equation}

With appropriate scaling in the infinite regions as $x\backsim\pm\log\kappa N$,
we find that the limit

\[
\lim_{N\rightarrow\infty}\kappa N\varphi(x-\log\kappa N)=-e^{x}
\]

and this allows to determine $E^{+}$ and $E^{-}$ as:

\begin{equation}
E^{\pm}=\sum_{i}e^{\pm\tilde{\gamma_{i}}^{\pm}}-\intop_{-\infty}^{\infty}\frac{dy}{2\pi}e^{\mp y}\log(1+e^{\epsilon^{\pm}(y)})
\end{equation}

where $\tilde{\gamma}^{\pm}$ is either $\tilde{\alpha}^{\pm}$ or
$\tilde{\beta_{j}}^{\pm}$, where $i$ runs over $\left\{ \pm,\ j=1,....,M\right\} .$

\section{Conclusion}

We analyzed a nontrivial relativistic integrable theory, namely the
boundary $\mathcal{M}(3;5)$ model, from the lattice point of view,
in the $(r=1,\ s=1)$ sector. This is a nontrivial non-unitary minimal
model, dual to the Lee-Yang model. The $A_{4}$ Restricted Solid-on-Solid
(RSOS) Forrester-Baxter model with trigonometric weights was solved
in the continuum scaling limit. We described the patterns of zeros
of the corresponding double row transfer matrix eigenvalues. Those
zeros are directly related to the RSOS paths on the lattice. Complementing
previous work done in the $\mathcal{M}(2,5)$ Lee-Yang model solved
before \cite{BDeebP,Deeb}, we adopted a similar approach to analyze
this model. However, the boundary conformal field theory (BCFT) model
is not fully solved, which prohibits direct comparison of the Virasoro
states with configurational paths, as well as the corresponding TBA
equations on the continuum side of the theory.

For the critical theory with integrable boundary, the transfer matrix
satisfies the same universal $Y$ system as \cite{Mercat}. The TBA
equations describe the finite-size scaling spectra of the $\mathcal{M}(3,5)$
model in the continuum scaling limit. The other sectors of this boundary
case are similar in their patterns of zeros with the only difference
is that some of them would contain a fixed zero at the center of the
analytic strip, while other sectors would not. 

The lattice description of the integrable scattering theory is used
in order to determine spectrum of the model. It may also allow further
to explore interesting and relevant physical quantities such as vacuum
expectation values and form factors. These quantities can be obtained
from the bootstrap approaches and they are available for the dual
Lee-Yang \cite{BajDeeb,hollo} but were not solved yet for the $\mathcal{M}(3,5)$
model. Future work should study the critical and massive $\mathcal{M}(3,5)$
models using the bootstrap methods and determine its scattering, reflection
and transmission matrices in different geometries, and its vacuum
expectation values as well as its form factors. It should also explore
other non-unitary models using both the RSOS lattice models and the
bootstrap approach.

\subsection*{Acknowledgments}

This work was performed during work at the Beirut Arab University
(BAU). I would like to thank the Physics department at BAU for its
support towards concluding this work. The advices of Maya Jalloul
on Latex code and figures were also beneficiary to finish the paper
in its current form.

\end{document}